\documentclass[
 reprint,
superscriptaddress,
 amsmath,amssymb,
 aps,
]{revtex4-2}
\usepackage{graphicx}
\usepackage{dcolumn}
\usepackage{bm}
\usepackage{siunitx}
\usepackage{hyperref}

\begin{document}

\preprint{APS/123-QED}

\title{Optical Phase Aberration Correction with an Ultracold Quantum Gas}

\author{Paul Hill}
\email{hill@physi.uni-heidelberg.de}
\affiliation{Physikalisches Institut der Universität Heidelberg, Heidelberg, Germany.}
\author{Philipp Lunt}
\affiliation{Physikalisches Institut der Universität Heidelberg, Heidelberg, Germany.}
\author{Johannes Reiter}
\affiliation{Physikalisches Institut der Universität Heidelberg, Heidelberg, Germany.}
\author{Maciej Ga{\l}ka}
\affiliation{Physikalisches Institut der Universität Heidelberg, Heidelberg, Germany.}
\author{Philipp M. Preiss}
\affiliation{Max Planck Institute of Quantum Optics, Garching, Germany.}
\author{Selim Jochim}
\affiliation{Physikalisches Institut der Universität Heidelberg, Heidelberg, Germany.}

\date{\today}
\begin{abstract}
We present an optical aberration correction technique for ultracold quantum gas experiments which directly utilizes the quantum gas as a wavefront sensor. The direct use of the quantum gas enables correcting aberrations that are otherwise impractical to measure, e.g. introduced by vacuum windows. We report a root-mean-square precision and accuracy of $0.01 \lambda$ and $0.03 \lambda$, respectively, and also show independently the reduction of aberrations through measurement of the trap frequency of our optical tweezer. These improvements were achieved for a tweezer size that is well below our imaging resolution. The present work is in particular intended to serve as a tutorial for experimentalists interested in implementing similar methods in their experiment.
\end{abstract}

\maketitle

\section{Introduction}
Over the last decades, significant progress has been achieved in the precise control and manipulation of complex many-body quantum systems down to the level of the individual constituting particles \cite{kaufman2021quantum, BruzewiczIons, bloch2005ultracold, gross2021quantum, superconducting, garcia2021semiconductor, slussarenko2019photonic}. This progress owes much to the use of optically engineered light fields, enabling advancements across various platforms. Prominent examples are found in cold ion or ultracold atom experiments, where accurately shaped light fields can be used for single-site addressing \cite{Nagerl, Wang, weitenberg2011single, labuhn2014single} or tailor-made optical potentials \cite{Anderson_1998, Barredo_2018, Gaunt_2013, Ritt, Henderson_2009, Wright_2013}.

To achieve the desired precision, the use of high-resolution optics is essential, requiring an optical setup to be free from any aberrations that could hinder performance at the diffraction limit. To address this issue, many experimental techniques and tools have been developed to measure and compensate for aberrations \cite{browaeys, bruce2015feedback, GAUTHIER20211}.  However, challenges arise when the targeted system is partially inaccessible to an experimentalist, i.e. when it is embedded in a solid-state environment or in a vacuum chamber. In such situations, aberrations typically can only be eliminated for certain parts of the setup.

Nevertheless, recent advancements have overcome these limitations by employing a variation of phase shift interferometry (PSI) \cite{Bruning:74, Cizmar2010} that directly uses the quantum system as a wavefront detector~\cite{Zupancic:16}. Originally developed for a neutral atom platform with a quantum gas microscope, this method has been successfully extended to cold ion quantum information processors \cite{shih_21}. By utilizing a digital micromirror device (DMD), aberrations in these systems were significantly reduced, enabling high-resolution single-site addressing of individual quantum particles.

In this work, we extend this approach to a bulk quantum gas consisting of a cold sample of $^6$Li atoms trapped in a coherent, optical double-well potential. The potential is created by coherently manipulating a Gaussian laser beam using a liquid-crystal spatial light modulator (SLM), thus rendering the double-well potential susceptible to aberrations present in the system. Imaging the atomic density is then sufficient to retrieve the relevant phase information which is required to cancel aberrations with the SLM. We benchmark our technique by injecting a known erroneous wavefront via the SLM and subsequently applying the aberration correction. We report a precision and accuracy of $0.01 \lambda$ and $0.03 \lambda$ (rms), respectively, while also observing a 4 \% increase in the trapping frequency of a tight optical tweezer generated with the SLM post-correction. We highlight that these improvements were achieved for a tweezer of about $1$ µm, despite our imaging resolution of the atomic density distribution being only $\sim 5$ µm. The method enables wavefront correction to the diffraction limit, even if calibration data can only be extracted at larger length scales. We emphasize that the methods for aberration correction described in the present paper are independent of the underlying experimental details and are applicable to a broad class of different designs.

The paper is organized as follows: Section \ref{sec:platform} provides a brief overview of the experimental setup and the requirements for applying the aberration detection scheme. In Section \ref{sec:methods}, we present a detailed description of the technique. Section \ref{sec:results} showcases the experimental results, and in Section \ref{sec:summary}, we summarize and discuss our findings.

\begin{figure}[t]
\includegraphics{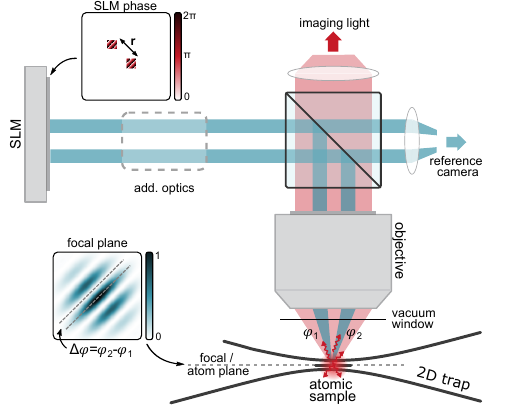}
\caption{\label{fig:setup} \textbf{Optical phase aberration correction.} The trapping light (blue) is reflected from the surface of the SLM, where its wavefront is altered by the phase pattern displayed on the SLM. We imprint a phase gradient on two small patches (upper inset) to separate the light reflected from these patches from the remaining light. This creates two small, parallel beams propagating through the optical setup. The main part of the light is focused through a high-resolution objective onto the atoms, while a minor fraction is focused on a reference camera. A vertical 2D trap confines the atoms to the focal plane of the objective. Resonant light (red) is used to image the atoms through the objective. As the two small beams created with the SLM traverse through different parts of the optical system, they acquire different phases $\varphi_1$, $\varphi_2$. Superimposing both beams in the focal plane of the objective forms an interference pattern (lower inset) which acts as a potential for the atomic sample. The phase difference $\Delta\varphi = \varphi_2 - \varphi_1$ both beams accumulate results in a shift of the position of the interference fringes (dashed lines), thereby providing a probe for detection of aberrations.}
\end{figure}

\section{\label{sec:platform}Experimental setup}
In Fig. \ref{fig:setup}, we outline the conceptual idea of the aberration correction and showcase the experimental setup \cite{Serwane_2011,lunt2024realization}. As crucial components to realize the correction scheme, we highlight the spatial light modulator (SLM) for the generation and manipulation of coherent optical beams, a two-dimensional (2D) confinement that restrains the atoms to the focal plane of the objective, and an imaging setup along the optical axis.
 
The SLM \footnote{here a liquid crystal SLM, Hamamatsu X10468-03 Liquid Crystal on Silicon (LCoS)} is placed in the Fourier plane of the atoms to create optical tweezers and other tailor-made optical potentials. We combine the SLM with a Fourier plane filter system (not depicted in Fig. \ref{fig:setup}) that allows us to single out light reflected from the SLM under certain angles. The unblocked light passes through an array of lenses and mirrors before it is projected onto the atoms via a final high-resolution objective ($\text{NA}=0.55$). 

In the present work \cite{Cizmar2010, Zupancic:16, shih_21}, we use the SLM to imprint a phase gradient onto two small patches of the outgoing wavefront (upper inset in Fig. \ref{fig:setup}). Together with the filter system, this generates two isolated optical beams overlapping in the atomic plane, while traversing the optical components on different paths. Aberrations present in the optical system will cause path-dependent phase shifts of the two beams. By observing these phase shifts through interference of both beams in the atomic plane we measure the aberrations present in the setup; an example of such an interference pattern is sketched in the lower inset of Fig. \ref{fig:setup}.

To confine the atoms to the focal plane of the objective we create a vertical 2D lattice by interfering two laser beams perpendicular to optical axis of the objective. The atoms are trapped in the combined potential of the two beams and the 2D confinement.

The atomic cloud is imaged $in$-$situ$ with resonant light via fluorescence imaging \cite{Bergschneider}. Given the strong correlation of the atomic cloud and the light intensity we effectively image the light field in the atomic plane (Appendix \ref{app:corr}).

The results given in the remainder of this paper are obtained for a Fermi degenerate sample of $\sim 100$ $^6$Li atoms \cite{Serwane_2011}. All trapping potentials are generated by a $\SI{1064}{nm}$ laser source. The objective is designed to be free of chromatic aberrations for the $\SI{1064}{nm}$ trapping light of the SLM and the $\SI{671}{nm}$ imaging light.

\section{\label{sec:methods}Wavefront detection technique}

When aberrations present in the optical setup are not too strong, the input field on the SLM and the field in the atom plane remain connected by an optical Fourier transform. Aberrations only enter the focal field by distorting the original wavefront of the incident beam. Combining the effect of the optics behind the SLM into a single lens with effective focal length $f$, we can write the field in the atomic plane as
\begin{equation}
    \left.E_\text{atoms}(\bm{x}) \sim\mathcal{F}\left[|E_\text{SLM}(\bm{x}')| e^{iW(\bm{x}')}\right](\bm{\nu})\right\vert_{\bm{\nu} = \bm{x}/\lambda f}, \label{eq:fourier}
\end{equation}
where $|E_\text{SLM}|$ is the amplitude of the electric field on the SLM, $W$ is a wavefront containing both the wavefront of the beam incident on the SLM and aberrations picked up during traversal of the non-ideal optical setup, and $\bm{\nu}$ denotes the coordinates in Fourier-space. By displaying a phase pattern according to $-W(\bm{x}')$ on the SLM a completely flat wavefront can be achieved, within the above approximation.

Following previously established methods \cite{Zupancic:16} we will now outline step-by-step how a coarse-grained version of the full wavefront $W$ can be obtained by locally probing relation Eq. \eqref{eq:fourier}. In section \ref{sec:phase_detection}, we first describe how phase differences between two patches of the wavefront can be extracted. Then, we present how shear maps of the wavefront can be recorded by moving a pair of patches across the SLM in section \ref{sec:wavefront}. Here, we also describe how a quasi-continuous wavefront is obtained from integration and interpolation of the recorded shear maps. 

\subsection{Detection of relative phase between wavefront patches}
\label{sec:phase_detection}
\begin{figure}[t]
\includegraphics{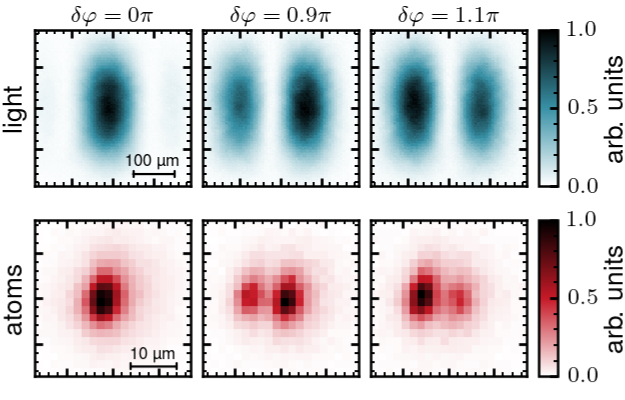}
\caption{\label{fig:psi_general} \textbf{Observation of phase-sensitive intensity pattern.} Images of the interference pattern directly recorded on the reference camera (top) and corresponding fluorescence images of the atomic density distribution (bottom). Here, the beams originate from two patches that are placed next to each other on the SLM, yielding a double-well pattern. By imprinting an additional phase $\delta\varphi$ onto one of the patches via the SLM we smoothly vary the position of the interference fringes. This is visible both in the light field on the reference camera and the atomic density. 
}
\end{figure}
\begin{figure*}[ht]
\includegraphics{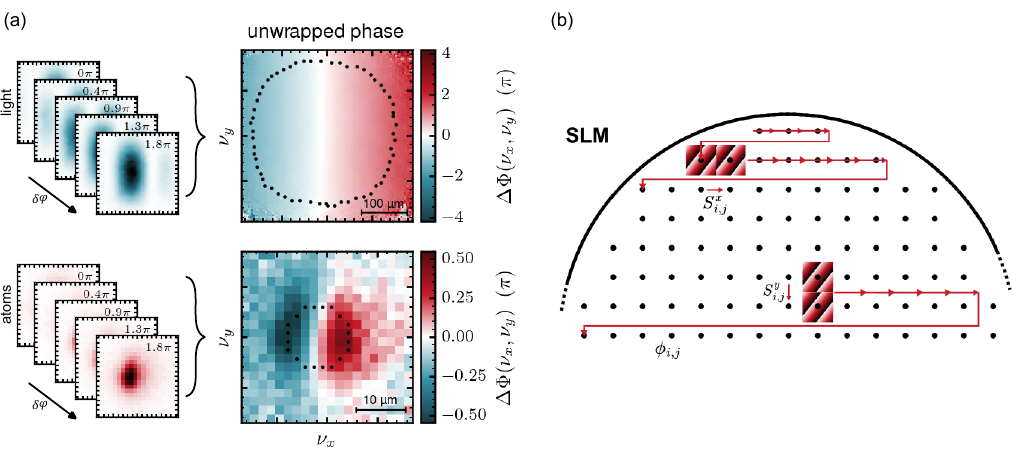}
\caption{\label{fig:psi_patch} \textbf{Phase shift interferometry and shear measurement scheme.} (a) Double-well pattern for different injected phases $\delta\varphi$ as directly recorded on the reference camera (c.f. Fig. \ref{fig:setup}), and via fluorescence imaging of the atoms, respectively (left column). The phase of the interference pattern is extracted via the PSI method and unwrapped for both the direct intensity measurement and the fluorescence image (right column). The dotted lines represent the outlines of the signal mask as described in the main text. (b) We display two patches centered on adjacent grid points of the grid $(x_i, y_j)$. By stacking them horizontally or vertically, we get access to the shears $S^x_{ij} = \varphi_{i+1, j} - \varphi_{i, j}$, and $ S^y_{ij} = \varphi_{i, j+1} - \varphi_{i, j}$, respectively. By moving both patches together across the SLM, as indicated by the red arrows, we measure the full shear maps. Horizontal, and vertical shear maps are obtained one after the other. The circle indicates the aperture of our objective.}
\end{figure*}

Using the SLM we imprint a phase gradient onto two square patches located at position $\bm{p}_1$ and $\bm{p}_2 = \bm{p}_1+\bm{r}$ (see Fig. \ref{fig:setup} inset) of the light reflected from the SLM. This allows us to crop out a small portion of the light field on the SLM, and hence also its wavefront. For small enough patch sizes $d$, it is reasonable to approximate the wavefront inside each patch to linear order as $W(\bm{x}) = \varphi_{\bm{p}_i} + \bm{g}_{\bm{p}_i}\cdot(\bm{x}-\bm{p}_i)$, where the phase offset $\varphi_{\bm{p}_i}$ (average wavefront in patch) and gradient of the wavefront $\bm{g}_{\bm{p}_i}$ is expected to vary between patches. For a field amplitude constant over both patches Eq. \eqref{eq:fourier} can then readily be integrated resulting in the intensity distribution
\begin{align}
    I(\bm{\nu}) ~&=~I_0(\bm{\nu})\left[1 + C(\bm{\nu})\cos(\Delta\Phi(\bm{\nu}))\right] \label{eq:int_pattern}\\
    \Delta\Phi(\bm{\nu}) &= ~2\pi \bm{r}\cdot\bm{\nu}+\Delta\varphi \label{eq:fringes}.
\end{align}
Here $I_0(\bm{\nu})$ denotes an envelope function, $C(\bm{\nu})$ encodes the contrast and the phase $\Delta\Phi(\bm{\nu})$ is defined in Eq. (\ref{eq:fringes}), which varies linear with $\bm{\nu}$. Eq. \eqref{eq:int_pattern} represents an interference pattern displaying characteristic fringes due to the oscillating cosine-term occurring with spatial frequency $|\bm{r}|$ set by the distance of the patches on the SLM. These fringes can be observed over a region determined by the sinc-shaped envelope $I_0(\bm{\nu})$, which is proportional to the inverse patch size $1/d$ in its extent. Their contrast $C(\bm{\nu})$ is unity only if the phase gradients inside the two patches don't differ, i.e. $\bm{g}_{\bm{p}_1} = \bm{g}_{\bm{p}_2}$. Most importantly, the phase difference $\Delta\varphi =  \varphi_{\bm{p}_2} -  \varphi_{\bm{p}_1}$ determines the position of the central interference fringe, thereby rendering the phase differences between different points of the wavefront $W$ observable, c.f. also the sketched intensity pattern Fig. \ref{fig:setup} (inset). Note that corrections to Eq. \eqref{eq:int_pattern}, \eqref{eq:fringes} arise if variations of the field amplitude are also considered (c.f. Appendix \ref{app:focal}).

In the configuration chosen here, in which both patches are located directly next to each other on the SLM, a double-well pattern with up to two interference fringes in the central maximum of the envelope is created. Then, the distance between the interference fringes is large enough such that we can resolve the fringe structure in the fluorescence signal of our atomic sample. Resulting images are presented in Fig. \ref{fig:psi_general}, showing both the atomic fluorescence signal and the light intensity taken with the reference camera (see Fig. \ref{fig:setup}). By adding an additional phase $\delta\varphi$ onto one of the beams, we deliberately shift the position of the interference maxima. A clear correlation of the atomic fluorescence signal with the light intensity images can be observed, and allows us to apply established techniques to retrieve the phase difference $\Delta\varphi$ from the atomic fluorescence data.

In principle fitting the recorded images with relation Eq. \eqref{eq:int_pattern} allows us to recover the relative phase offset $\Delta\varphi$ between both patches. However, here we use phase shift interferometry (PSI)~\cite{Bruning:74, Cizmar2010} instead. The PSI method allows us to reconstruct the entire optical phase $\Delta\Phi(\bm{\nu})$ from any signal of the form $I^*(\bm{\nu})=f(I(\bm{\nu}))$ for an arbitrary, unknown function $f$ that can be well approximated by a finite power expansion (for derivation see Appendix \ref{app:psi}). Specifically, this means that $\Delta\Phi(\bm{\nu})$ can faithfully be extracted from the atomic density of a Fermi gas if a local density approximation is assumed (for discussion see Appendix \ref{app:corr}). The phase offset $\Delta\varphi$ is then readily obtained via a linear fit to the observed phase data following Eq.~\eqref{eq:fringes}.

To perform PSI we deliberately add another phase shift $\delta\varphi$ to the patch at $\bm{p}_1$. Scanning the phase shift over $2\pi$, according to $\delta\varphi_n = 2\pi n/N, n=0,...,N-1$, yields a sequence of patterns $I^*_n(\bm{\nu})$, that is periodic in $n$. This allows us to directly extract the phase (for details see Appendix~\ref{app:psi})
\begin{equation}
    \Delta\Phi(\bm{\nu}) = \arctan\left(\frac{\sum_{n=0}^{N-1} I^*_n(\bm{\nu}) \sin\delta\varphi_n}{\sum_{n=0}^{N-1} I^*_n(\bm{\nu}) \cos\delta\varphi_n}\right). \label{eq:psi}
\end{equation}
In Fig.~\ref{fig:psi_patch}a we show measurements of the direct light intensity, and the corresponding atomic fluorescence signal for different values of $\delta\varphi$. From each sequence of images, we extract the relative phase $\Delta\Phi(\bm{\nu})$ using Eq. \eqref{eq:psi}. The resulting unwrapped phases are presented in Fig. \ref{fig:psi_patch}a.

We select the signal region (in which the phase is well measured and follows the linear relationship expected from Eq. \eqref{eq:fringes}) by constructing a simple mask based on a threshold criterion for the averaged intensity data, $\sum_n I^*_n > I^*_{th}$. While the expected linear relationship from Eq. \eqref{eq:fringes} can be observed over a large region on the reference image, the atomic-fluorescence images contain more noise and we find significant non-linear deviations at the outskirts of the interference pattern. This requires us to choose a tight threshold for the signal mask, which results in a smaller signal region (dashed line in Fig.~\ref{fig:psi_patch}a). To fit the phase profile, we choose a robust linear RANSAC model \cite{fischler_bolles_1981}, which actively rejects outliers not following the expected linear relationship. In this way, we minimize the effects of noise and non-linearities, increasing the robustness of the fitting routine. The fitted parameters then give us an estimate for the relative phase offset $\Delta\varphi$. We attribute the presence of non-linearities to the fact that the atomic fluorescence signal is in general not a point-wise function of the intensity distribution, but rather a function of the intensity pattern as a whole. This effect is expected to be especially prominent in small atomic systems for which a local density approximation is no longer justified (Appendix \ref{app:corr}).

\subsection{Wavefront recovery \label{sec:wavefront}}
To reconstruct the entire wavefront we measure the relative phase between patches for varying locations across the SLM. Here, multiple schemes exist \cite{Cizmar2010, Zupancic:16, shih_21}. By keeping one patch fixed as a reference and moving the second patch across the SLM we can directly measure the wavefront relative to the reference patch. The varying distance between both patches leads to an altering number of fringes in the central maximum of the interference pattern, i.e. to a changing fringe density. This, in turn, requires sufficient resolution of the imaging system. While this typically does not pose a problem for direct measurements on a camera, our imaging setup for the atoms only allows us to resolve a single fringe, or double-well, in the interference pattern, hence rendering a direct wavefront measurement impossible. To circumvent this problem, we measure the relative phase between two patches stacked vertically and horizontally, respectively, as we now move both patches across the SLM, as illustrated in Fig. \ref{fig:psi_patch}b. This allows us to record shear maps of the wavefront.

Let $\varphi_{ij}$ denote the phase offset $\varphi_{\bm{p}}$ of a patch located on a grid $M\times M$ - grid, i.e. $\bm{p} = (x_i, y_i) = (x_0, y_0) + (id, jd)$ ($i, j = 0, ..., M-1$). The shear scheme described above measures
\begin{equation}
\begin{split}
    S^x_{ij} &= \varphi_{i+1, j} - \varphi_{i, j},\\
    S^y_{ij} &= \varphi_{i, j+1} - \varphi_{i, j}, \label{eq:shears}
\end{split}
\end{equation}
where the shear map $S^x$ is sampled on a $M-1\times M$ grid and $S^y$ on a $M\times M-1$ grid.\\
We remark that the finite size of the patches restricts our ability to measure the erroneous wavefront $W(x, y)$ for high spatial frequencies. Furthermore, if phase and amplitude gradients within one patch are present simultaneously, the relative phase offset between two patches retrieved with the PSI-method differs slightly from $\Delta\varphi$, as discussed in Appendix \ref{app:focal}. However, iterative measurement and correction of the erroneous wavefront should eliminate this discrepancy since the wavefront is flattened successively and phase gradients are diminished. For both the direct and the shear scheme, gradients present in the initial wavefront must not be too strong, as all phases can only be determined modulo $\pi$. Hence, the phase difference between adjacent grid points should not exceed $\pm\pi/2$, i.e. $|S^{x,y}_{ij}|<\pi/2$.

Eq. \eqref{eq:shears} in one dimension is straightforward to invert as shears can be simply added up to yield the phase at site $(i, j)$. In the two-dimensional case, the problem becomes over-constrained since two points are connected by a multitude of distinct paths along which shears can be added up to yield the phase at the endpoint. This results in further constraints for the shear maps, namely that shears along closed paths have to add up to zero. In the presence of measurement errors, these constraints will most certainly be violated making it a non-trivial task to determine the phase, i.e. to properly integrate the shear maps. To overcome this problem we can first find the integrable shear maps that are closest to the measured shear maps, i.e. we project the measured shear maps onto the subspace of integrable shear maps. Once in the integrable subspace, the wavefront can be obtained directly via integration similar to the one-dimensional case. Note that projecting onto the integrable subspace also functions as a measurement error mitigation strategy. The whole procedure is equivalent to minimizing the least squares problem
\begin{equation}
\label{eq: least squares}
\begin{split}
    f(\{\phi_{ij}\}) &= \sum_{ij}(\phi_{i+1, j} - \phi_{i, j} - S^x_{ij})^2\\
    &\quad+\sum_{ij}(\phi_{i, j+1} - \phi_{i, j}-S^y_{ij})^2
\end{split}
\end{equation}
with respect to the phase map $\phi_{ij}$ (with constraint $\phi_{00}=0$), already proposed in \cite{Talmi:06}. This directly provides us with a prediction of the discretized wavefront $\varphi_{ij}$ up to a constant.

To obtain a phase pattern in the full resolution of our SLM we interpolate the integrated phase field using biharmonic splines \cite{PolySplines}. Spline interpolation has the advantage that local errors only influence the interpolation result locally, which is in contrast to global interpolation with e.g. Zernike polynomials. On the downside, especially local statistical errors, are not smoothed out as much.

\section{\label{sec:results}Results}

\subsection{Flattening the wavefront}
\begin{figure}
\includegraphics{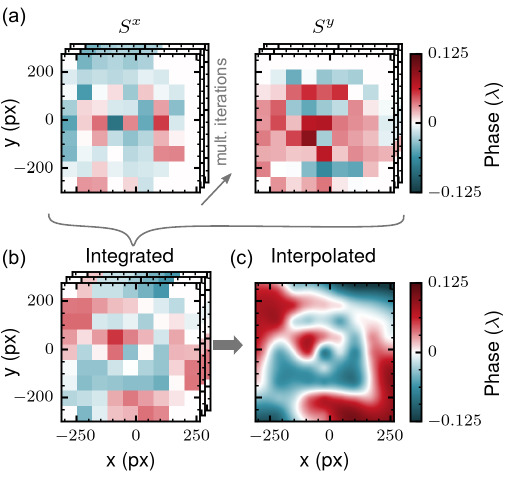}
\caption{\label{fig:flattening the wavefront} \textbf{Reconstruction of the aberrated wavefront}. (a) By measuring the relative phase $\Delta \varphi$ between adjacent sampling grid points $x_i$ and $y_j$, the shears in both $x$ and $y$ direction are obtained and can be combined in the respective shear maps $S^x$ and $S^y$. (b) Integrating the shear maps yields the discrete phase map $\phi_{ij}$ where corner points containing no signal have been neglected. (c) The integrated phase map is interpolated using biharmonic splines to obtain a smooth version, whose inverse is displayed on the SLM to compensate for the measured wavefront error. The resulting map shown here is the combined correction map after three subsequent iterations of the algorithm. The aperture of our objective corresponds to a diameter of $630$ px on the SLM.}
\end{figure}
We first cancel aberrations introduced by the optical components located before our objective, which allows us to directly use the reference camera (c.f. Fig. \ref{fig:setup}). As the wavefront curvature is initially quite strong we resort to recording a Shack-Hartmann correction pattern on our camera\cite{Lopez-Quesada:09}. This is done by moving one patch across the SLM and recording the shift of the resulting focus on the reference camera, giving a measurement of the gradient of the wavefront $\nabla W$. The wavefront is obtained by fitting the gradient field of Zernike polynomials to $\nabla W$. We then apply the method described in section \ref{sec:methods} to obtain a more accurate measurement of the residual aberrations. In total, we find deviations from a flat wavefront of about $\sigma^0_\text{ptv}\sim2\lambda$, which we partially attribute to the non-perfect collimation of the beam incident on the SLM. After iterative application of our correction method, we find residual aberrations of about $\sigma^0_\text{ptv}=0.02\lambda$.

Next, we turn to measure the aberrations introduced by the remaining optical elements directly on the atoms; primarily stemming from the objective and the vacuum window. To this end, we prepare a degenerate Fermi gas of $^6$Li in an optical dipole trap (ODT) at 685G, below the Fesbach resonance of Lithium (hyperfine $|1 \rangle$ and $|3 \rangle$ states) \cite{Zuern_2013}. At 660G we load the atoms from the ODT in the double-well potential created with the SLM. After switching off the ODT we ramp on the vertical 2D confinement and perform a final evaporation ramp of the double-well potential. This reduces the chemical potential and temperature further, allowing us to resolve the fringes of the double-well pattern. Finally, the atoms are imaged with resonant light to retrieve their density distribution. The respective fluorescence images are recorded for six different injected phases $\delta \varphi$ for each position $(x_i,y_j)$ of the patch on the $9\times 9$ sampling grid with a patch size of $70$ pixels (for comparision, the SLM has a 800x600 px display with a pixel size of $20$ µm). Following the procedure described in section \ref{sec:phase_detection}, the relative phase $\Delta \varphi$ between adjacent grid sites is calculated. By subsequently moving the patches across the SLM the entire wavefront is sampled and the relative phases are combined into the shear maps $S^x$ and $S^y$ shown in Fig. \ref{fig:flattening the wavefront}a for the respective scanning direction $x$ and $y$. As opposed to taking a single fluorescence image per setting $(x_i,y_i,\delta \varphi)$ we instead record at least 20 raw images and combine them into an averaged image to increase the signal-to-noise ratio of the phase extraction. Due to the circular aperture of the optical system (corresponding to a diameter of $\sim 630$ px on the SLM), no atoms are trapped in the corners of the quadratic sampling grid and the recovered shears are neglected. We integrate the two shear maps by solving the least squares problem in Eq. \eqref{eq: least squares} and obtain the discrete phase map $\phi_{ij}$ shown in Fig. \ref{fig:flattening the wavefront}b. To obtain a smooth phase map we interpolate the discrete phase map using biharmonic splines and project its inverse on the SLM to compensate for the measured aberrations after subtracting a constant offset phase and linear gradient. 

After the first iteration, the residual aberrations still exceed the later referred precision, thus we repeat the algorithm. In each iteration, a new interpolated phase map is displayed on the SLM in addition to the previous correction maps to subsequently reduce the residual wavefront error. After three iterations the residual error remains constant indicating the convergence of the algorithm. 

The phase map in Fig. \ref{fig:flattening the wavefront}c represents the combined interpolated phase maps derived from the three iterations which is used to compensate for the full measurable wavefront error by projecting its inverse on the SLM. 
Since we first ran the algorithm on a reference camera (c. f. Fig. \ref{fig:setup}) before running it on the atoms, the measured wavefront error on the atoms mainly stems from the objective and the vacuum window. Moreover, we attribute the circular feature in the center of the phase map to the point mirror (aluminium, 1 mm diameter) on top of our objective which is used to create a Magneto-Optical Trap (MOT).
Quantitatively, the measured wavefront error corresponds to a peak-to-valley (ptv) deviation of $\sigma_\text{ptv}=0.25\lambda$ and a root-mean-square deviation from a perfectly flat wavefront of $\sigma_\text{rms}=0.05\lambda$ expressed in units of the wavelength $\lambda$. In addition, the residual wavefront error after three iterations of the algorithm allows us to specify a precision of $\sigma_\text{ptv}=0.08\lambda$ and $\sigma_\text{rms}=0.01\lambda$, respectively. Corner points that are clipped by the circular aperture of the objective and hence do not yield an adequate signal for phase reconstruction have been neglected in the calculation of the ptv and rms deviations.

Assuming that the recovered phase map yields a flat wavefront, we now proceed to characterize the performance of the algorithm for wavefront reconstruction.

\begin{figure}
\includegraphics{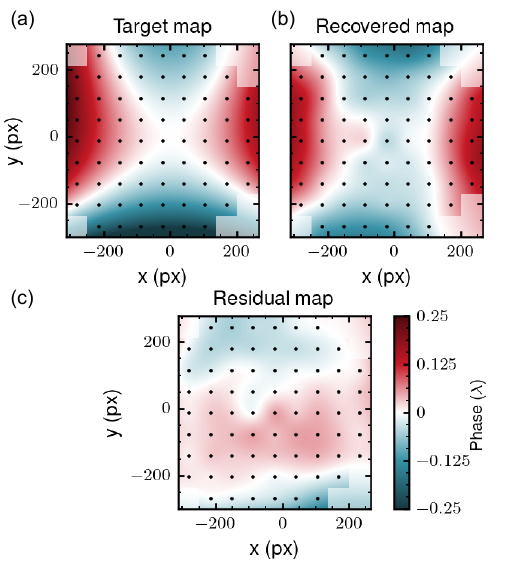}
\caption{\label{fig:benchmark} \textbf{Benchmarking the algorithm}.(a) A target phase map comprising a combination of astigmatism and coma is projected on the SLM to artificially introduce well-known aberrations in the optical system allowing for precise bench-marking. (b) Running the algorithm on the atoms yields a combined recovered phase map which is used to compensate for the artificially introduced aberrations. (c) The residual phase map shows a wavefront error of $\sigma_\text{ptv}=0.20\lambda$ and $\sigma_\text{rms}=0.03\lambda$ confirming the high wavefront reconstruction accuracy enabled by the algorithm. The quantitative analysis has only been performed on the unshaded area indicated by the black dots where a clear signal was achieved on the atoms. The aperture of our objective corresponds to a diameter of $630$ px on the SLM.}
\end{figure}

\subsection{Benchmarking the algorithm}
Not only does the SLM provide the necessary tool to measure and compensate for wavefront errors, but it can also be used to introduce artificial aberrations. While artificial optical aberrations are generally undesired in experimental applications they enable us to benchmark the performance of the wavefront correction algorithm on the atoms. For this purpose, we project the target phase map shown in Fig. \ref{fig:benchmark}a on the SLM and try to subsequently recover it by performing the algorithm on the atoms. The target wavefront is chosen to represent a typical optical aberration as it might be present in a real optical setup comprising a combination of astigmatism and coma with a peak-to-valley strength of approximately $0.6\lambda$. In three subsequent iterations, we measure the shears, reconstruct the integrated wavefront, and project the inverse of the interpolated wavefront on the SLM.

Fig. \ref{fig:benchmark}b shows the combined reconstructed phase map from the three iterations after which the algorithm converged. Albeit minor deviations from the target map the reconstructed phase map exhibits a high level of congruence with the target map and hence indicates the high accuracy that can be achieved with the algorithm. Quantitatively, we find a residual error on the sampled region of $\sigma_\text{ptv}=0.20\lambda$ and $\sigma_\text{rms}=0.03\lambda$, respectively.

\subsection{Alternative measure: Trap frequencies}
We have shown that we can recover the injected phase map using the aberration correction algorithm. However, since our method might still be biased towards some residual error that is not accessible to the algorithm, we measure the axial trapping frequency of a tight optical tweezer pre- and post-correction to verify the absolute reduction of phase aberrations.

By illuminating almost the whole aperture of our objective we generate a single, tightly focused optical tweezer with a waist of roughly $1$ µm in the atom plane. We deterministically prepare two fermions (hyperfine $|1 \rangle$ and $|3 \rangle$ states) in the ground state of the tweezer by employing a spilling technique~\cite{Serwane_2011}, which uses an axial magnetic field gradient to control the number of bound states in the trap. The magnetic Feshbach resonance allows us to tune the interactions such that state $|1 \rangle$ and $|3 \rangle$ do not interact with each other.

\begin{figure}
\includegraphics{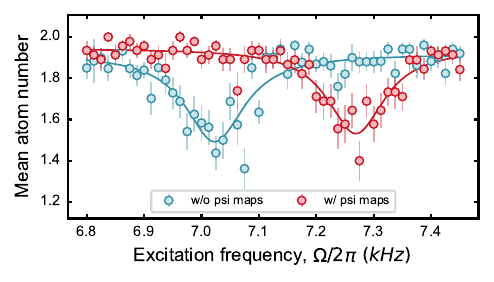}
\caption{\label{fig:trap_freq} \textbf{Alternative verification of the aberration correction: trap frequencies}. We verify the reduction of phase aberrations independently of the previous benchmarking method by measuring the trap frequencies of two non-interacting atoms in the ground state of a tightly focused optical tweezer with and without the correction phase map (PSI map). We observe an increase in the axial trap frequency post-correction (red curve).}
\end{figure}

We measure the axial trap frequency of the tweezer by modulating the power at a frequency $\Omega$ for $50$ cycles with a peak-to-peak amplitude of $4 \%$. On resonance, atoms are excited to higher trap levels which are removed subsequently by a second spilling procedure. We count the remaining atoms in the ground state by releasing the atoms from the tweezer into a magneto-optical trap and measuring their fluorescence signal \cite{Hu_1994, Serwane_2011}. Due to the even spatial symmetry of the modulation, only an even multiple of the harmonic oscillator trap frequency can be excited (parity conservation). The observed resonance spectra with and without the measured correction map are shown in Fig. \ref{fig:trap_freq}. The measured excitation frequency without correction is $\Omega _\text{w/o} = 7023 \pm 6 \text{Hz}$. In the corrected tweezer the frequency is raised to $\Omega _\text{w/} = 7269 \pm 4 \text{Hz}$, corresponding to an increase in trap frequency of $ (\Omega _\text{w/} - \Omega _\text{w/o}) / \Omega _\text{w/o} \approx 4\%$.

\section{\label{sec:summary}Summary and Discussion }
We demonstrate aberration correction for high-precision generation of tailor-made optical potentials using a spatial light. Specifically, we focus on using our cold atomic sample to measure the aberrations introduced by the last optical components in our setup, our high-NA objective, and the vacuum window of the science chamber. The aberrations introduced by other components in the setup were removed by running our algorithm on a reference camera prior to the measurements described here. We find that the residual aberrations have a peak-to-valley variation of $\sigma_\text{ptv}=0.25\lambda$ and a root-mean-square error of $\sigma_\text{rms}=0.05\lambda$. When applying these corrections to our tight optical tweezer, we observe an increase in trapping frequency of about $4\%$. We emphasize that these results represent improvements on a length scale undercutting the resolution of our imaging system. 

To further benchmark the algorithm we deliberately inject a known erroneous wavefront into the system via the SLM. By comparing it to the wavefront retrieved with our correction method we find a peak-to-value deviation of $\sigma_\text{ptv}=0.20\lambda$, while the root-mean-square deviation is $\sigma_\text{rms}=0.03\lambda$. Furthermore, we specify a precision of $\sigma_\text{ptv}=0.08\lambda$ and $\sigma_\text{rms}=0.01\lambda$ based on residual aberrations of the wavefront measured in the last iteration of the algorithm. We conclude that the aberrations introduced by our high-NA objective and vacuum window are on the order of our measurement accuracy. Indeed most of the aberrations in the whole setup are introduced by preceding components, with a peak-to-valley variation of about $\sigma^0_\text{ptv}\sim2\lambda$. Note however, that we don't fully illuminate our objective and hence don't probe the outer most region of its aperture, which is typically where the largest aberrations are introduced.

By correcting for both measured wavefronts with the SLM we obtain an optical system with minimal aberrations, rendering the creation of precise coherent optical potentials possible. As one limitation for the performance of the correction method outlined in this work we see the optical power available for the generation of the double-well potential. Here, more power would allow for smaller patch sizes to be used, such that the wavefront can be sampled more finely. As available optical power on the SLM typically drops with the distance from the optical axis, more power would also allow for a larger region of the wavefront to be explored.\\

\paragraph*{Funding}
This work has been supported by the Heidelberg Center for Quantum Dynamics, the DFG Collaborative Research Centre SFB 1225 (ISOQUANT), Germany’s Excellence Strategy EXC2181/1-390900948 (Heidelberg Excellence Cluster STRUCTURES) and the European Union’s Horizon 2020 research and innovation program under grant agreements No.~817482 (PASQuanS),  No.~725636 (ERC QuStA) and No. 948240 (ERC UniRand). This work has been partially financed by the Baden-Württemberg Stiftung. 

\appendix

\section{Field in the Focal Plane\label{app:focal}}
We present some more details about sec. \ref{sec:phase_detection}. Our main concern is to evaluate the Fourier transform of the light field on the SLM for a single patch, i.e. to compute Eq. \eqref{eq:fourier} with the amplitude and phase
\begin{align*}
\begin{split}
    |E_\text{SLM}(\bm{x})| &= m(\bm{x}-\bm{p})\left[a_{\bm{p}} + \bm{u}_{\bm{p}}\cdot(\bm{x}-\bm{p})\right]\\
    W(\bm{x}) &= \varphi_{\bm{p}} + \bm{g_p}\cdot (\bm{x} - \bm{p}).
\end{split}
\end{align*}
Here $m(\bm{x}) = \Theta(d/2-|x|) \Theta(d/2-|y|)$ denotes the patch mask expressed in terms of the Heaviside- $\Theta$ function and we allow for variations of the field amplitude and phase to linear order. $a_{\bm{p}}$ denotes the average amplitude in the patch, while $\bm{u}_{\bm{p}}$ encodes an amplitude gradient. $\varphi_{\bm{p}}$ and $\bm{g_p}$ are the phase offset and phase gradient as defined in the main text. Computing the Fourier transform then yields
\begin{equation}
\begin{split}
   \hat{E}(\bm{\nu}) &= \int d^2\bm{x}|E_\text{SLM}(\bm{x})| e^{iW(\bm{x})}e^{2\pi i \bm{\nu}\cdot\bm{x}}\\
    &= e^{i\varphi_{\bm{p}} + 2\pi i\bm{\nu}\cdot\bm{p}}\int d^2\bm{x} m(\bm{x})(a_{\bm{p}} + \bm{u}_{\bm{p}}\cdot\bm{x}) e^{ i(2\pi\bm{\nu}+\bm{g}_{\bm{p}})\cdot\bm{x}}\\
                      &\equiv e^{i\varphi_{\bm{p}} + 2\pi i\bm{\nu}\cdot\bm{p}} F(\bm{\nu}+\bm{g}_{\bm{p}}/2\pi) \label{eq:e_focal_acc}.
\end{split}
\end{equation}
The integral $F(\bm{\nu})$ further evaluates to
\begin{equation}
\begin{split}
    F(\bm{\nu}) = d^2a_{\bm{p}}j_0(\xi_x)j_0(\xi_y)\left[1 + i\frac{d\bm{u}_{\bm{p}}}{2a_p} \cdot 
    \begin{pmatrix}
           j_1(\xi_x)/j_0(\xi_x) \\
           j_1(\xi_y)/j_0(\xi_y)
    \end{pmatrix}\right]\label{eq:e_focal_f}
\end{split}
\end{equation}
where $j_n(x)$ denote the spherical Bessel functions of the first kind, $\xi_x = \pi d\nu_x$ and $\xi_y = \pi d\nu_y$. For the flat amplitude case, $\bm{u}_{\bm{p}}=0$, one obtains the usual sinc-envelope, with amplitude $|F(\bm{\nu})| = d^2a_{\bm{p}}|j_0(\xi_x)j_0(\xi_y)|$. Corrections for $\bm{u}_{\bm{p}}\neq 0$ to first order are linear in $d\bm{u}_{\bm{p}} / a_{\bm{p}}$ for the phase of $F(\bm{\nu})$ and quadratic for the amplitude. Close to the central peak ($\xi_x,\xi_y\ll 1$) one may approximate
\begin{equation}
\begin{split}
    |F(\bm{\nu})| &\equiv A_{\bm{p}}(\bm{\nu}) \approx d^2a_{\bm{p}}|j_0(\xi_x)j_0(\xi_y)|\\
    \arg F(\bm{\nu}) &\approx 2\pi\frac{d^2}{12a_{\bm{p}}}\bm{u}_{\bm{p}}\cdot\bm{\nu}\equiv 2\pi\hat{\bm{u}}_{\bm{p}}\cdot\bm{\nu} \label{eq:e_focal_arg},
\end{split}    
\end{equation}
where the latter corresponds to a phase gradient. Applying these approximations to Eq. \eqref{eq:e_focal_acc} yields
\begin{equation}
    \hat{E}(\bm{\nu}) \approx A_{\bm{p}}(\bm{\nu} + \bm{g}_{\bm{p}} / 2\pi)\exp\left[2\pi i \bm{\nu}\cdot \hat{\bm{g}}_{\bm{p}} + i\phi_{\bm{p}}\right],\label{eq:e_focal}
\end{equation}
with $\hat{\bm{g}}_{\bm{p}} = \bm{p} + \hat{\bm{u}}_{\bm{p}}$ and $\phi_{\bm{p}} = \varphi_{\bm{p}} + \bm{g}_{\bm{p}}\cdot\hat{\bm{u}}_{\bm{p}}$. In general, amplitude gradients will result in deviations most prominently occurring around the nodes of the ideal sinc-shaped envelope, i.e. of $j_0(\xi_{x,y})$, due to the diverging terms in Eq. \eqref{eq:e_focal_f}. There, sharp features like kinks in amplitude, and jumps in the phase, are smoothed out by higher-order corrections. Note that due to the sign of the sinc-function, an additional phase factor $e^{i\pi}$ may appear in Eq. \eqref{eq:e_focal} even in the ideal case if it is evaluated outside the central maximum of the envelope $|F(\bm{\nu})|$.\\

Combining two beams originating from distinct patches $\bm{p}_1$ and $\bm{p}_2$ will result in the interference pattern given by Eq. \eqref{eq:int_pattern} but a different phase
\begin{equation}
\begin{split}
    I(\bm{\nu}) ~&=~I_0(\bm{\nu})\left[1 + C(\bm{\nu})\cos(\Delta\Phi(\bm{\nu}))\right]\\
    \Delta\Phi(\bm{\nu}) &= ~2\pi (\hat{\bm{g}}_{\bm{p}_2}-\hat{\bm{g}}_{\bm{p}_1})\cdot\bm{\nu}+\phi_{\bm{p}_2}-\phi_{\bm{p}_1}
\end{split}
\end{equation}
as compared to Eq. \eqref{eq:fringes}. The contrast and envelope are given by
\begin{equation}
\begin{split}
C(\bm{\nu}) &= \frac{2A_{\bm{p}_1}(\bm{\nu} + \bm{g}_{\bm{p}_1} / 2\pi)A_{\bm{p}_1}(\bm{\nu} + \bm{g}_{\bm{p}_2} / 2\pi)}{I_0(\bm{\nu})}\\
I_0(\bm{\nu}) &= |A_{\bm{p}_1}(\bm{\nu} + \bm{g}_{\bm{p}_1} / 2\pi)|^2 + |A_{\bm{p}_2}(\bm{\nu} + \bm{g}_{\bm{p}_2} / 2\pi)|^2.\label{eq:contrast_env}
\end{split}
\end{equation}
In general, a difference in the gradients $\Delta\bm{g}=\bm{g}_{\bm{p}_2}-\bm{g}_{\bm{p}_1}$ will lower the contrast and hence decrease the signal from which the phase of the pattern can be extracted. Note that even in the presence of amplitude gradients the relative phase $\Delta\phi=\phi_{\bm{p}_2}-\phi_{\bm{p}_1} = \bm{g}_{\bm{p}_2}\cdot\hat{\bm{u}}_{\bm{p}_2} - \bm{g}_{\bm{p}_1}\cdot\hat{\bm{u}}_{\bm{p}_1} + \Delta\varphi$ approaches $\Delta\varphi$ in the limit of small gradients $\bm{g}_{\bm{p}}$ of the wavefront. Hence, the true phase difference $\Delta\varphi$ can still be found by iterative application of the correction algorithm, which should diminish existing phase gradients in each iteration.

\section{Phase Shift Interferometry\label{app:psi}}
As explained in the main text a phase $\delta\varphi_n = 2\pi n/N, n=0,...,N-1$, imprinted onto the patch at $\bm{p}_1$ yields a sequence of patterns $I^*_n(\bm{\nu})$, that is periodic in $n$. Here $I^*_n(\bm{\nu}) = f(I_n(\bm{\nu}))$ describes any signal that derives from the intensity pattern Eq. \eqref{eq:int_pattern} in a point-wise sense. Hence at any point $\bm{\nu}$ one can set
\begin{equation}
     I^*_n(\bm{\nu}) \equiv f(a_0 + a_1\cos(2\pi n/N - \Phi)), \label{eq:In}
\end{equation}
with $a_0 = I_0(\bm{\nu}), a_1 = I_0(\bm{\nu})C(\bm{\nu})$ and $\Phi = \Phi(\bm{\nu})$. \\
If $f$ allows for a (absolutely convergent) power expansion $f(x)=\sum_kf_kx^k$, Eq. \eqref{eq:In} can be recast into
\begin{align*}
\begin{split}
    I^*_n &= \sum_{k=0}^\infty f_k \sum_{l=0}^k {k \choose l}a_0^{k-l}\left[a_1\cos(2\pi n/N - \Phi)\right]^l\\
    &= \sum_{k=0}^\infty f_k \sum_{l=0}^k {k \choose l}a_0^{k-l}\left(\frac{a_1}{2}\right)^l\sum_{m=0}^l{l\choose m}e^{i(\frac{2\pi n}{N}-\Phi)(2m-l)}.
\end{split}
\end{align*}
Reordering for orders of the oscillating exponential gives
\begin{align*}
\begin{split}
    I^*_n &= \sum_{j=-\infty}^{\infty}e^{i(\frac{2\pi n}{N}-\Phi)j}\sum_{k=|j|}^\infty f_k\sum_{\substack{l=|j|,\\|j|+2, ...}}^k {k \choose l} a_0^{k-l}\left(\frac{a_1}{2}\right)^l{l\choose \frac{j+l}{2}}\\
    &\equiv \sum_{j=-\infty}^{\infty}e^{i(\frac{2\pi n}{N}-\Phi)j} \hat{b}_{|j|},
\end{split}
\end{align*}
where the symbols $\hat{b}_{j}$ were defined and ${l\choose (j+l)/2} = {l\choose (-j+l)/2}$ was used. Factoring out the unique terms that oscillate with $n$ yields
\begin{align*}
\begin{split}
    I^*_n&= \sum_{j=0}^{N-1}\left[e^{i(\frac{2\pi n}{N}-\Phi)j}\sum_{m=0}^\infty \epsilon_{j,m}e^{-iNm\Phi}\hat{b}_{j+Nm}\right] + c.c.\\
    &\equiv \frac{1}{2}\sum_{j=0}^{N-1}e^{i(\frac{2\pi n}{N}-\Phi)j} b_j(\Phi) + c.c\\
    &= \sum_{j=0}^{N-1}|b_j(\Phi)| \cos\left(\frac{2\pi jn}{N}-j\Phi + \arg b_j(\Phi)\right),\\
\end{split}
\end{align*}
where $\epsilon_{j, m} = 1-\frac{1}{2}\delta_{j,m}$, in terms of the usual Kronecker delta and the symbols ${b}_{j}(\Phi)$ where defined. The oscillating cosine terms are still not linearly independent and the sum needs to be reduced further to
\begin{widetext}
\begin{align*}
\begin{split}
    I^*_n&= \Re(b_0(\Phi))+\sum_{j=1}^{\lfloor\frac{N}{2}\rfloor}\epsilon_{j,\frac{N}{2}}\bigg\{|b_j(\Phi)| \cos\left(\frac{2\pi jn}{N}-j\Phi + \arg b_j(\Phi)\right)+|b_{N-j}(\Phi)| \cos\left(\frac{2\pi jn}{N}+(N-j)\Phi - \arg b_{N-j}(\Phi)\right)\bigg\},\\
    &= \Re(b_0(\Phi))+\sum_{j=1}^{\lfloor\frac{N}{2}\rfloor}\epsilon_{j,\frac{N}{2}}\bigg\{\cos\left(\frac{2\pi jn}{N}\right)\bigg[|b_j(\Phi)|\cos\left(j\Phi - \arg b_j(\Phi)\right)+|b_{N-j}(\Phi)|\cos((N-j)\Phi - \arg b_{N-j}(\Phi))\bigg]\\
    &\quad+\sin\left(\frac{2\pi jn}{N}\right)\bigg[|b_j(\Phi)|\sin\left(j\Phi - \arg b_j(\Phi)\right)-|b_{N-j}(\Phi)|\sin((N-j)\Phi - \arg b_{N-j}(\Phi))\bigg]\bigg\},
\end{split}
\end{align*}
\end{widetext}
which now represents a discrete Fourier series expansion of $I^*$ in terms of cosines and sines. Focusing in particular on the first harmonic terms ($j=1$) we note that the coefficients simplify considerably in the case of $b_{N-1}=0$. This situation is realized if the expansion of $f$ can be truncated at order $N-2$ ($f_k\approx0, k>N-2$). One finds
\begin{equation}
    b_1 = 2\sum_{k=1}^{N-2} f_k \sum_{l=1, 3, ...}^k {k \choose l}a_0^{k-l}\left(\frac{a_1}{2}\right)^l{l\choose \frac{l+1}{2}} \in \mathbb{R},
\end{equation}
which is real and independent of $\Phi$. Under such assumptions, the phase $\Phi$ we are interested in can be extracted from the first harmonic contribution of the Fourier expansion. The coefficients $b_1\cos(\Phi)$ and $b_1\sin(\Phi)$ are then given through
\begin{align*}
\begin{split}
     b_1\cos(\Phi) &= \frac{2}{N}\sum_{n=0}^{N-1} I^*_n(\bm{\nu}) \cos(2\pi n/N) \\
     b_1\sin(\Phi) &= \frac{2}{N}\sum_{n=0}^{N-1} I^*_n(\bm{\nu}) \sin(2\pi n/N).
\end{split}
\end{align*}
Combining both expressions directly yields the phase of the interference pattern, Eq. \eqref{eq:psi} given in the main text.

\section{Light intensity and atomic fluorescence signal\label{app:corr}}
The exact relationship between the light intensity, defining the potential $V(\bm{x})$ in which the atoms are confined, and the atomic density distribution $n(\bm{x})$ is not known. Only in simple situations, an analytic expression can be given. Assuming a local density approximation the external potential can be absorbed into the chemical potential $\mu\rightarrow \mu - V(\bm{x})$ and the atomic density takes the form 
\begin{equation}
n(\bm{x}) = f(\mu(\bm{x}), T, a), \label{eq:dens_gas}
\end{equation}
where $f$ is the free-space solution for temperature $T$ and inter-particle scattering length $a$. Under such assumptions the considerations of appendix \ref{app:psi} apply and the PSI method can be used provided that the injected phase $\delta\varphi$ is sampled finely enough. Note, that in the particular cases of a Bose-Einstein-condensate with weak repulsive interactions or a degenerate 2D Fermi gas Eq. \eqref{eq:dens_gas} considerably simplifies to $n(\bm{x}) \sim \mu(\bm{x})$ \cite{bec_superfluidity}.

In our case of a cold, mesoscopic gas of Feshbach-molecules the above approximations are not well justified. The atomic density at point $\bm{x}$ will depend then also on the potential at positions $\bm{x'}$ other than $\bm{x}$. Hence, we expect phases extracted using the PSI method to deviate from the ideal case. Such deviations are expected to be especially prominent in regions of low density. Similar effects are introduced by imperfections in the imaging process, e.g. random walk of the atoms during fluorescence imaging and the finite resolution of the optical imaging system.

\end{document}